# Convective Heat Transfer and Pressure Drop Characteristics of Graphene-Water Nanofluids in Transitional Flow

**Çayan Demirkır, Hakan Ertürk[1]**

Bogazici University

Department of Mechanical Engineering

Istanbul, 34342, Turkey

## ABSTRACT

The convective heat transfer and flow behavior of graphene-water nanofluids are studied experimentally by focusing on transitional flow. Graphene-water nanofluids with different particle mass fractions (0.025, 0.1 and 0.2%) are produced following two-step method and using PVP as a surfactant. Thermo-physical characterization is performed by measuring viscosity and thermal conductivity of the nanofluids. Convection characteristics are experimentally studied from laminar to turbulent flow regimes. It is seen that pressure drop increases dramatically in the transition region, and laminar to turbulent transition shifts to lower Reynolds numbers with increasing nanoparticle concentration. The transition initiates at a Reynolds number of 2475 for water, while it initiates at 2315 for the nanofluid with 0.2% particle mass fraction. Increase in mean heat transfer coefficient and Nusselt numbers are nearly identical at different Reynolds numbers and axial positions along the test tube in the laminar flow for nanofluids and water due to dominance of conduction enhancement mechanisms on the heat transfer increase in laminar flow. Beyond laminar flow regime, enhancement of Nusselt number is observed indicating that thermophoresis and Brownian motion are more effective heat transfer augmentation mechanisms. The maximum heat transfer enhancement is observed as 36% for a Reynolds number of 3950.

[1] Corresponding author: H. Ertürk (hakan.erturk@boun.edu.tr)

**Nomenclature**

| | | | |
|---|---|---|---|
| $c_p$ | specific heat capacity | $T_w(x)$ | local wall temperature |
| $d$ | particle diameter | $u_m$ | mean fluid velocity |
| $D$ | hydraulic diameter of pipe | $x$ | axial distance from heated part of the test section |
| $f$ | friction factor | $x^*$ | dimensionless axial distance |
| $\bar{h}_D$ | mean convective heat transfer coefficient | | |
| $h_x$ | local convective heat transfer coefficient | *Greek Symbols* | |
| $k$ | thermal conductivity coefficient | $\Delta P$ | pressure drop along the test section |
| $k_B$ | Boltzmann constant | | |
| $L$ | length of the test tube | $\phi$ | particle mass concentration |
| $\dot{m}$ | mass flow rate | $\varphi$ | particle volume concentration |
| $Nu_x$ | local Nusselt number | $\mu$ | dynamic viscosity |
| $\overline{Nu}_D$ | mean Nusselt number | $\pi$ | pi number |
| $P$ | fluid pressure | $\rho$ | density |
| $Pr$ | Prandtl number | $\sigma$ | experimental uncertainty |
| $q''$ | heat flux applied to test section | *Subscripts* | |
| $q_f$ | heat transferred to system from heater | $bf$ | base fluid |
| $q_{in}$ | power supplied to heater | $i$ | inlet |
| $Re_D$ | Reynolds number | $nf$ | nanofluid |
| $\bar{T}_f$ | average of the inlet and outlet temperatures | $np$ | nanoparticle |
| $\bar{T}_w$ | average temperature of the thermocouples mounted on pipe | $o$ | outlet |
| $T_m$ | fluid mean temperature | | |
| $T_m(x)$ | local mean fluid temperature | | |

## 1. Introduction

There has been an increasing effort for miniaturization of systems for many different applications necessitating an increase in heat transfer performance for safe, reliable operation and sustaining or increasing efficiency. However, there exist performance limitations due to material properties. One way to tackle this problem is designing and developing advanced materials with improved thermal transport characteristics. Adding solid particles into common heat transfer fluids like water, certain glycols such as ethylene glycol (EG) and propylene glycol (PG), is known to augment heat transfer properties. Using micro or millimeter-sized particles leads to agglomeration, sedimentation, clogging problems, erosion of the pipelines, and high-pressure drop [1,2]. With the developments of nanotechnology, the use of nanometer-sized particles led to introduction of colloidal suspension of nano-particles that is also known as nanofluids [3]. Nanofluid preparation methods, their thermo-physical and optical properties have been investigated for many different materials [4–7] and it was observed that different mechanisms contribute to the observed superior heat transfer properties. Brownian motion, nano-layering of liquid molecules around particles, percolation through clustering nanoparticles and thermophoresis or Soret effect are argued to be the most effective enhancement mechanisms of heat transport [8,9]. Numerous theoretical and empirical models are formulated to accurately estimate the thermal conductivity of nanofluids by considering these mechanisms [10].

Forced convection and flow behavior of nanofluids have also been an area of interest for researchers. While many researchers observed significant heat transfer enhancement for different nanoparticles [11,12], whereas some reported little or no enhancement [13–15]. In addition to the metallic and ceramic nanoparticles, carbon-based nanoparticles have been gaining attention for producing nanofluids. Wang et al. [16] studied laminar flow of multi-walled carbon nanotube (MWCNT)-water nanofluids in a horizontal circular pipe, and showed remarkable increase in heat transfer up to 190% even at low Reynolds numbers such as 120. They also observed a linear increase in the pressure drop with Reynolds number. Hemmat Esfe et al. [17] investigated turbulent flow of up to 1% COOH-functionalized MWCNT-water nanofluids for particle volume fractions. They reported that average heat transfer coefficient increase is 78%, and Nusselt number increase is 36% for the Reynolds numbers varying between 5000 and 27,000. Baby and Ramaprabhu [18] prepared nanofluids by dispersing functionalized hydrogen exfoliated graphene (f-HEG) nanoparticles into EG/water mixture and studied turbulent heat convection. They measured entrance and developed heat transfer enhancements separately and showed that heat transfer increases up to 170% at the entrance, whereas the increase is around 140% at outlet of the pipe for 0.01% particle volume fraction. Ghozatloo et al. [19] studied graphene-water nanofluids focusing on the thermal characterization under laminar regime. The effect of temperature and concentration is investigated and remarkable heat transfer augmentation is reported with rising temperature or concentration. They observed 15% heat transfer augmentation at 25ºC as the particle mass fraction is increased from 0.025 to 0.1% , whereas enhancement is 24% at 38ºC showing that the change in temperature is more effective than changing particle concentration. Akhavan-Zanjani et al. [20] studied laminar forced convection of graphene-water nanofluids along a uniformly heated annular tube and reported a heat transfer enhancement of about 14% for 0.02% concentration at a Reynolds number of 1850. Selvam et al. [21] prepared graphene-water/EG nanofluids for 0.1-0.5% particle volume concentrations, and conducted a research on convective heat transfer characteristics over a broad range of Reynolds number. They observed up to 170% and 96% increase in heat transfer coefficient and Nusselt number, respectively, at a Reynolds number of 6790. Yarmand et al. [22] studied the functionalized graphene nano-platelet-water nanofluid, and reported 19% and 26% enhancement for heat transfer coefficient and Nusselt number, respectively, with 9% increase in friction factor for a Reynolds number of 17500, with 0.1% particle mass concentration.

Many studies focus on laminar or turbulent convective heat transfer, whereas the number of studies related to internal transitional flow is very limited. This is due to the fact that thermal engineers and equipment manufacturers prefer to operate systems either in laminar or turbulent regimes. However, it is critical to reveal the transitional behavior, where sudden changes can be observed. Considering that the presence of particles has different effects on flow and heat transfer behavior, their effect on transitional flow must also be identified. Hence, there is a need for experimental investigations of transitional behavior of different heat transfer fluids. Meyer et al. [23] studied heat transfer and pressure drop of the MWCNT-water nanofluids for particle concentrations from 0.33 to 1.0% by focusing on the transitional flow regime. They observed no significant difference in heat transfer for laminar regime, whereas up to 33% enhancement in turbulent regime is reported. Moreover, transition starts at lower Reynolds numbers when compared to water. Negligible pressure drop increase is seen in the turbulent regime, whereas increase is higher for laminar flow. Chandrasekar et al. [24] assessed thermal and hydrodynamic behavior of 0.1% by volume $Al_2O_3$-water nanofluids subject to constant heat flux under transition region. They reported that maximum Nusselt number enhancement is 34% for a Reynolds number of 5000. Moreover, they did not observe a significant change of pressure drop with respect to that of pure water. Wusiman et al. [25] studied Cu-water nanofluids with four different particle concentrations and in a wide range of Reynolds number (300-16,000). They showed that heat transfer increased with concentration under laminar regime, but it decreased under transitional flow for most of the concentrations. Whereas, the increase in heat transfer can be up to 25% depending on the Reynolds number and concentration for turbulent flow. Cabaleiro et al. [26] prepared ZnO-EG/water nanofluids to investigate the heat transfer performance for a flow range varying from laminar to turbulent at various particle concentrations. They applied different heat fluxes on the nanofluids with particle mass fraction of 1.0%, and their results indicated no significant heat transfer augmentation for the prepared nanofluids with respect to water under transitional flow regime.

Even though there are limited number of studies regarding the transitional flow behavior of some metallic and ceramic nanoparticles, such as $Al_2O_3$, ZnO and Cu, there exists no studies in the literature that investigated the transitional flow of graphene dispersed nanofluids. Therefore, an experimental study on the convective heat transfer and pressure drop behaviors of the graphene-water nanofluids is carried out, focusing on laminar to turbulent transition considering graphene-water nanofluids with particle mass concentrations up to 0.2%. Effect of particle concentration on the onset of transition is investigated based on heat transfer and pressure drop measurements for the first time.

## 2. Experimental set-up and procedure

### 2.1 Preparation and characterization of nanofluids

In this study, graphene-water nanofluids with particle mass fractions of 0.025%, 0.1% and 0.2% are prepared by using two-step method using distilled water as a base fluid. Graphene nano-platelets used in this study have 99% purity, 5-10 nm thickness, 5-10 μm lateral size according to the report of the manufacturer (Grafen Chemical Industries, Turkey). Graphene is a hydrophobic material, and it does not dissolve in the polar solvents such as water; therefore, surface-active material, PVP K30, is used to prepare stable nanofluids. PVP is weighed by precision balance (Kern PFB, ±10 mg) in a ratio of 1:1 with respect to graphene mass and added to 400 ml de-ionized (DI) water. The mixture is stirred by a mechanical mixer (Heidolph, RZR 2021) for 15 minutes at 1600 rpm. Graphene nanoparticles are then added into the solution and the mixture is stirred for 45 minutes more at the same speed. The suspension is put into a water bath and ultrasonically mixed (Hielscher UP400S, using sonotrode H40) for two hours by applying 200 W. A circulating chiller (PolyScience 9106A12E) is used to adjust the temperature of water bath to 8°C.

Although morphological, thermal and rheological characterization of the prepared nanofluids are comprehensively reported in our previous study [27], some important points are expressed here for the completeness of discussion. Environmental scanning electron microscopy (ESEM) imaging is performed (Philips XL30 ESEM-FEG/EDAX) before the mixing processes in order to verify the manufacturer's report regarding the size of the dry nanoparticles (Fig. 1a).

It is observed that the manufacturer report is consistent with ESEM image for most of the particles, whereas there are some agglomerations that size up to 5μm. Following the preparation of the nanofluids, scanning transmission electron microscopy (STEM) imaging is carried out using the same instrument as a part of morphological characterization. STEM image of the graphene-water nanofluids is presented in Fig. 1b, where it can be seen that most of the agglomerations observed in Fig. 1a are broken and almost all nanoparticles' size is under 1μm as a result of the ultrasonic mixing process. Further stability analysis of the graphene-water nanofluids is carried by measuring the zeta potential. The measured zeta potential of the nanofluids is about -40 mV indicating that the suspensions produced are stable. The measurements are performed right after the nanofluids' preparation, whereas a long term stability analysis can be found in [27].

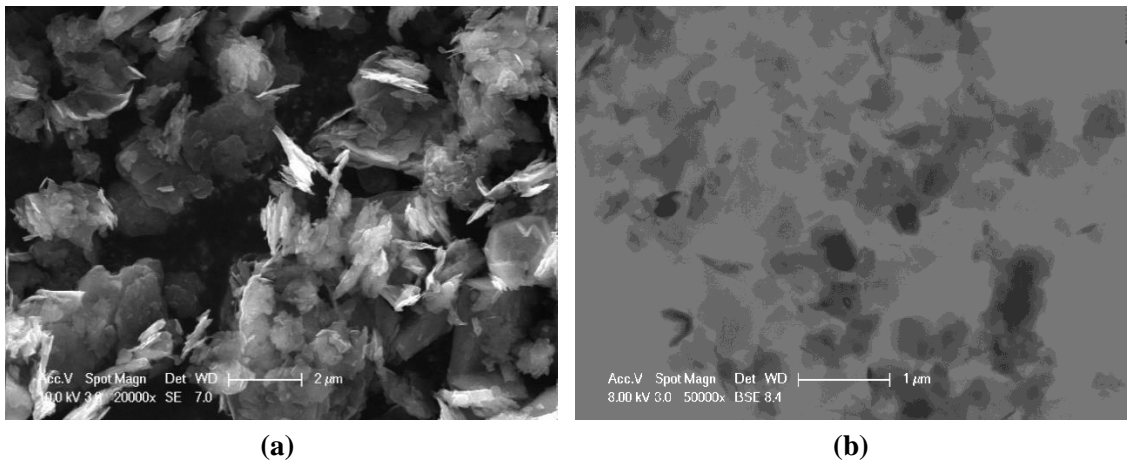

(a)                                              (b)

**Fig. 1.** (a) ESEM image of graphene nano-platelets before mixing, (b) STEM image of graphene-water nanofluid (0.1% by particle mass fraction)

Thermal characterization is carried out by measuring thermal conductivity of prepared graphene-water nanofluids using thermal conductivity analyzer (Decagon KD2 Pro, ±5%). Measurements are performed at 25ºC and average of 10 measurements is reported ensuring repeatability. Viscosity is measured by a cone-plate rheometer (Brookfield DV-III Ultra, ±1% of full-scale range). A recirculating chiller (PolyScience, 9106A12E) is used to stabilize the temperature of the nanofluid sample during measurements. Thermal conductivity and viscosity measurements are validated using water and EG. Relative thermal conductivity and viscosity of prepared nanofluids with respect to water are shown in Fig. 2.

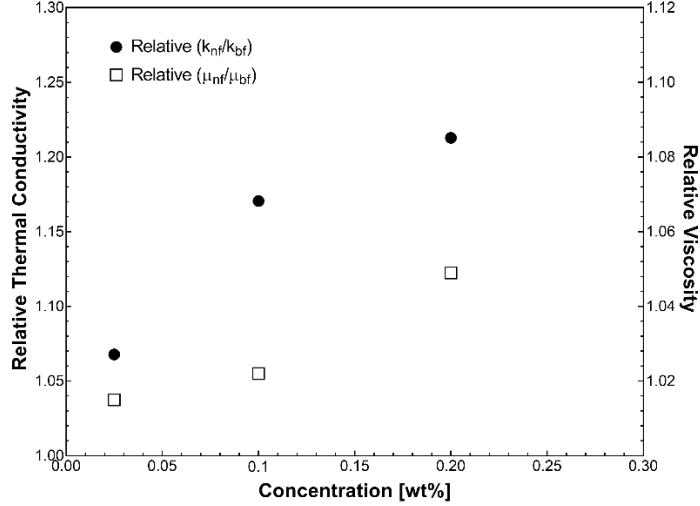

**Fig. 2.** Relative thermal conductivity and viscosity of nanofluids with 0.025, 0.1 and 0.2% mass concentration

Density ($\rho$) and specific heat ($c_p$) of the nanofluid defined in terms of the particle mass fraction, $\phi$, can be determined according to basic mixture theory as

$$\rho_{nf} = \frac{\rho_{np}\rho_{bf}}{(1-\phi)\rho_{np} + \phi\rho_{bf}} \tag{1}$$

$$c_{p,nf} = \phi c_{p,np} + (1-\phi)c_{p,bf} \tag{2}$$

## 2.2 Experimental set-up

The experimental test set-up is designed and assembled to investigate internal forced convection from laminar to turbulent regimes. As seen in Fig. 3, pumped fluid flows through the horizontal test tube, which is comprised of a 2.1 m long circular copper pipe with 6 mm inner diameter and 1 mm wall thickness. The first 0.6 m of the pipe is not heated so that the flow develops hydrodynamically before it enters the heated section. The latter part of the pipe is heated by a coiled nichrome heater wire to obtain uniform heat flux. The assembly procedure of the test set-up is carried out as follows.

Firstly, the entire copper pipe is covered by fiberglass sleeve to ensure electrical insulation. Heater wire is then wrapped around the last 1.5 m of the tube to establish uniform heat flux throughout the temperature measurements. After the sleeve is shaved in determined axial locations, thermocouples are attached to the outer pipe wall for measuring the local surface temperatures in the heated section. Then, heater wire is covered by zinc phosphate-based cement for obtaining a homogenous heat distribution along the pipe. Heat and electricity resistant fiberglass insulation tape and fireproof cloth tape are wrapped on the heated section respectively to prevent any ignition at high temperatures. Finally, two-layered glass wool is covered for minimizing the heat loss from the system. Heater wire is connected to AC power source at both ends and is supplied with 400 W using a potentiometer.

The fluid leaving the copper tube enters the concentric heat exchanger, where water circulated by the chiller (Polystat 12920) reduces the temperature of the fluid coming from the test section and stabilizes the temperature before entering the storage tank. The fluid is then pumped by a centrifugal pump (Iwaki RD-20). Flow rate is controlled by a valve, and it is measured by a turbine flowmeter (Sea, YF-S402, ±2% of reading). The flowmeter was calibrated before the very first experiment and checked before each experiment. Pressure drop is measured by using two pressure transducers (Setra C206, ±0.13% of full scales that are 0-25 and 0-50 PSIG, respectively) that are mounted to the valves placed at the inlet and outlet of the copper pipe, and they are also connected to the data acquisition unit to record the pressure values. Validation and calibration of pressure transducers are carried out by measuring static water pressure for different heights.

T-type thermocouples with special error limits (Omega Inc., ±0.5°C) are used for measuring the temperatures at different axial locations on the copper pipe. Nine thermocouples are mounted on the pipe at axial positions ($x/D$) of 10, 33.3, 46.6, 60, 80, 100, 120, 200, 240 (x starting from the heating section) and two thermocouples are positioned in the plastic pipe, at the inlet and outlet of the test tube to measure the bulk mean fluid temperature. These last two thermocouples are installed after the flow meter and the valve to which the outlet pressure transducer is installed, respectively. The flow is mixed at both locations so that the mean temperature can be correctly measured. Temperature measurements are taken one hour after the system started circulating to ensure that steady state has been reached. The measurements are collected by a data acquisition unit (Agilent 34970A) for 10 minutes, and the mean value is processed to avoid any fluctuations in the data that appear in transitional and turbulent flows. Calibration of the thermocouples is carried out by using a constant temperature water bath. Pressure drop and convective heat transfer experiments are performed simultaneously. Measurements are repeated at least two times and the mean value is reported if the results are consistent with each other. The test setup is cleaned by circulating DI water after each nanofluid experiment.

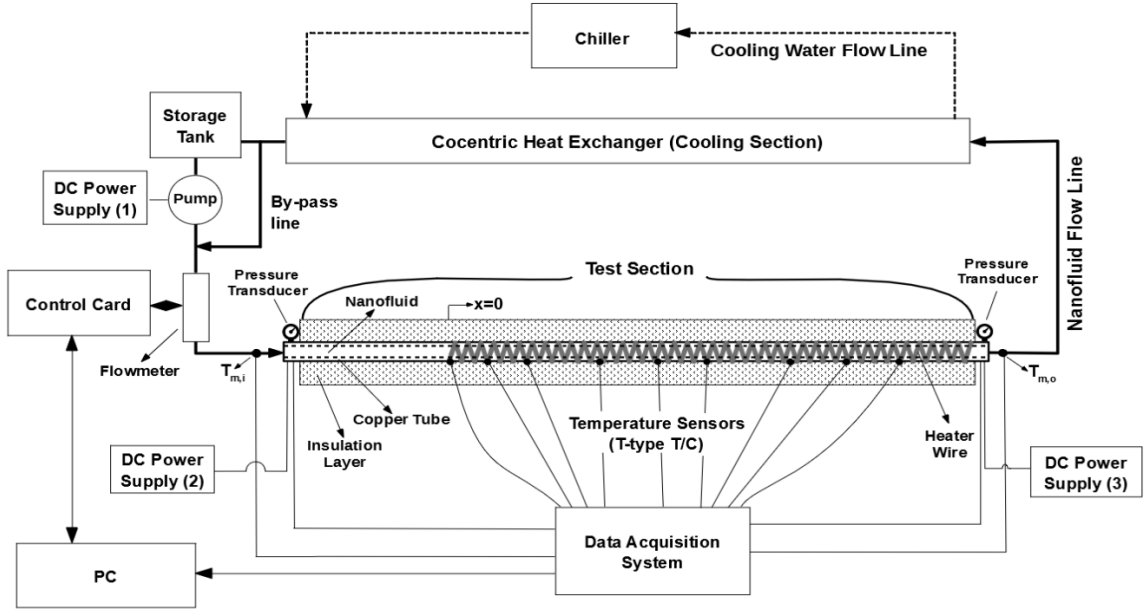

**Fig. 3.** Experimental system used in this study

## 2.3 Data Reduction

The equations used in the analysis of pressure drop and heat transfer measurements are shown as follows. Some well-known correlations are used to validate the accuracy of the experimental set-up under laminar and turbulent flows for water, and determine the transition region.

### *Pressure Drop and Friction Factor*

Pressure transducers are mounted on the inlet and outlet of the test section to measure the pressure drop ($\Delta P$) along the system. For laminar flow, measured pressure drop is validated using Hagen-Poiseuille equation.

$$\Delta P = P_i - P_o = \frac{32 \mu u_m L}{D^2} \tag{3}$$

where $P_i$ and $P_o$ are the fluid pressures at the inlet and outlet of the test tube; $\mu$, $u_m$, $L$ and $D$ represent dynamic viscosity (Pa), mean fluid velocity (m/s), tube length (m) and tube diameter (m), respectively.

Then, friction factor is calculated by Darcy-Weisbach formula.

$$f = \frac{\Delta P \left(\frac{D}{L}\right)}{\frac{1}{2}\rho u_m^2} \tag{4}$$

Poiseuille correlation (Eq. 5) is used to validate the friction factor under laminar flow, whereas Blasius (Eq. 6) and Petukhov (Eq. 7) correlations are used beyond laminar region.

$$f = \frac{64}{Re_D} \tag{5}$$

$$f = 0.316 Re_D^{0.25} \tag{6}$$

$$f = (1.82 \log Re_D - 1.64)^{-2} \tag{7}$$

*Heat Transfer*

Convective heat transfer measurements are first performed by using DI water in the test system, then for the nanofluids with the mass fractions of 0.025%, 0.1% and 0.2%. The insulation quality of the test setup is determined by comparing the power supplied to the heater ($q_{in}$) with the heat transferred to system from heater ($q_f$). While $q_{in}$ is set to 400 W in this study, $q_f$ is calculated as follows

$$q_f = \dot{m} c_p (T_{m,o} - T_{m,i}) \tag{8}$$

where $\dot{m}$ is the mass flow rate (kg/s), $T_{m,o}$ and $T_{m,i}$ are the fluid outlet and inlet mean temperatures (°C), respectively that are measured by submerging a thermocouple into the flow at the inlet of the test pipe as explained in the previous section.

Heat loss of the experimental system is obtained by using following equation and it is found to be lower than 8%.

$$q_{loss} = \left(1 - \frac{q_f}{q_{in}}\right) x\ 100 \tag{9}$$

Local convective heat transfer coefficient, $h_x$, is calculated based on Newton's law of cooling;

$$h_x = \frac{\frac{q_f}{\pi DL}}{T_w(x) - T_m(x)} \tag{10}$$

$T_w(x)$ and $T_m(x)$ represent local wall and mean fluid temperatures (°C) at axial position $x$, respectively. The mean temperature at a position $x$ can be defined as;

$$T_m(x) = T_{m,i} + \frac{\left(\frac{q_f}{L}\right)x}{\dot{m}c_p} \tag{11}$$

where and $x$ is axial distance from the heated part of the pipe (m). Temperature difference between inner and outer wall of the copper pipe is less than 0.1 °C, which is well within the measurement uncertainty.

Local Nusselt number, $Nu_x$, can be determined after calculating the local heat transfer coefficient.

$$Nu_x = \frac{h_x D}{k} \tag{12}$$

where $k$ is thermal conductivity (W/m.K).

Among many empirical equations are developed to estimate the local Nusselt number under laminar flow conditions, Shah and London [28] correlation is used to validate the test setup for laminar flow in this study.

$$Nu_x = \begin{cases} 1.302(x^*)^{-\frac{1}{3}} - 1, & x^* \leq 0.0005 \\ 1.302(x^*)^{-1/3} - 0.5, & 0.00005 \leq x^* \leq 0.0015 \\ 4.364 + 0.263(x^*)^{-0.506}\exp(-41x^*), & x^* > 0.0015 \end{cases} \tag{13}$$

where $x^* = (x/D)/(Re_D Pr)$ and represents dimensionless axial distance.

A flow range from laminar to turbulent is considered in this study. In addition to laminar flow, it is necessary to represent the equations used for the transitional and turbulent flows. Due to the instabilities in these regions, mean values of the heat transfer coefficient and Nusselt number are used instead of the local values. The mean heat transfer coefficient, $\bar{h}_D$, for the entire test unit is defined as;

$$\bar{h}_D = \frac{\frac{q_f}{\pi D L}}{\bar{T}_w - \bar{T}_f} \tag{14}$$

Here, $\bar{T}_w$ is the average temperature of the thermocouples mounted on the test unit and $\bar{T}_f$ is average of the inlet and outlet temperatures. The mean Nusselt number, $\overline{Nu}_D$, is defined based on mean heat transfer coefficient accordingly.

$$\overline{Nu}_D = \frac{\bar{h}_D D}{k} \tag{15}$$

Mean Nusselt number calculated by Gnielinski correlation for laminar flow is used for the validation study that is carried out with DI water [29].

$$\overline{Nu}_D = \left[ \overline{Nu}_{D,1}^3 + 0.6^3 + \left(\overline{Nu}_{D,2} - 0.6\right)^3 + \overline{Nu}_{D,3}^3 \right]^{1/3} \tag{16}$$

with

$$\overline{Nu}_{D,1} = 4.354,$$

$$\overline{Nu}_{D,2} = 1.953 \sqrt[3]{Re_D \Pr \left(\frac{D}{L}\right)},$$

$$\overline{Nu}_{D,3} = 0.924 \sqrt[3]{\Pr} \sqrt{Re_D \left(\frac{D}{L}\right)},$$

Similarly, Gnielinski correlation is used to validate the mean Nusselt number of water for turbulent flow [30]

$$\overline{Nu}_D = \frac{(f/8)(Re_D - 1000)(Pr)}{1 + 12.7 \sqrt{\frac{f}{8}} (Pr^{2/3} - 1)} \qquad \begin{array}{c} 2300 \leq Re_D \leq 10^6, \\ 0.5 < Pr < 2000 \end{array} \tag{17}$$

where $2300 \leq Re_D \leq 10^6$ and $0.5 < Pr < 2000$, and the friction factor, $f$, is estimated by using Petukhov equation.

## 2.4 Uncertainty Analysis

Uncertainties in flow rate, temperature, thermal conductivity and heat flux measurements lead to uncertainty in the measured heat transfer coefficient, Nusselt number, and friction factor. Therefore, an uncertainty analysis for local heat transfer coefficient, Nusselt number and friction factor is carried out [31]. In the calculations, the uncertainties of the measuring devices presented in the previous section are used.

$$\sigma_{h_x} = \left[\left(\frac{\partial h_x}{\partial \dot{m}}\sigma_{\dot{m}}\right)^2 + \left(\frac{\partial h_x}{\partial T_w}\sigma_T\right)^2 + \left(\frac{\partial h_x}{\partial T_{m,i}}\sigma_T\right)^2 + \left(\frac{\partial h_x}{\partial q"}\sigma_{q"}\right)^2\right]^{1/2} \quad (18)$$

The uncertainty of the heat flux, $\sigma_{q"}$, is calculated by Eq. 19 and 20 as

$$q" = \frac{\dot{m}c_p}{\pi D L}(T_{m,o} - T_{m,i}) \quad (19)$$

$$\sigma_{q"} = \left[\left(\frac{\partial q"}{\partial \dot{m}}\sigma_{\dot{m}}\right)^2 + \left(\frac{\partial q"}{\partial T_{m,o}}\sigma_T\right)^2 + \left(\frac{\partial q"}{\partial T_{m,i}}\sigma_T\right)^2\right]^{1/2} \quad (20)$$

Then, uncertainty of the Nusselt number is found by following equation.

$$\sigma_{Nu_x} = \left[\left(\frac{\partial Nu_x}{\partial \dot{m}}\sigma_{\dot{m}}\right)^2 + \left(\frac{\partial Nu_x}{\partial T_w}\sigma_T\right)^2 + \left(\frac{\partial Nu_x}{\partial T_{m,i}}\sigma_T\right)^2 + \left(\frac{\partial Nu_x}{\partial q"}\sigma_{q"}\right)^2 + \left(\frac{\partial Nu_x}{\partial k}\sigma_k\right)^2\right]^{1/2} \quad (21)$$

In the Fig. 4, the experimental uncertainties for the $h_x$ and $Nu_x$ under laminar water flow are shown to the 8% and 9%, respectively. Beyond the laminar region, the uncertainties for the $\bar{h}_D$ and $\overline{Nu}_D$ are calculated similarly and increase up to 10% and 11% respectively.

In addition to heat transfer, the uncertainty calculations for the pressure drop and friction factor are performed by equations below.

$$\sigma_{\Delta P} = \left[\left(\frac{\partial \Delta P}{\partial P1}\sigma_{P1}\right)^2 + \left(\frac{\partial \Delta P}{\partial P2}\sigma_{P2}\right)^2\right]^{1/2} \quad (22)$$

$$\sigma_f = \left[\left(\frac{\partial f}{\partial \Delta P}\sigma_{\Delta P}\right)^2 + \left(\frac{\partial f}{\partial u_m}\sigma_{u_m}\right)^2\right]^{1/2} \quad (23)$$

It is shown that the uncertainty of the pressure drop measurements for the range from laminar to turbulent flow is below 2%, whereas friction factor is below 5%.

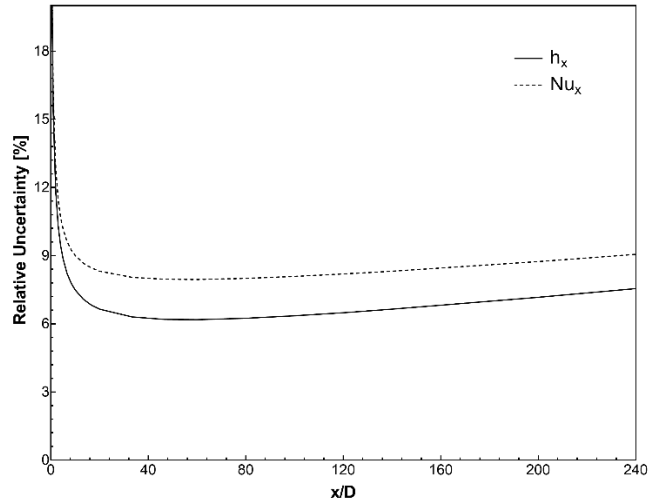

**Fig. 4.** Relative uncertainty of $h_x$ and $Nu_x$ along the test tube

### 3- Results and Discussion

**3.1- Pressure Drop and Friction Factor**

Pressure drop (ΔP) and friction factor (f) of graphene-water nanofluids with different particle concentrations (0.025, 0.1 and 0.2%) are investigated for a Reynolds number range of 1400 to 4000. A validation study is carried out with DI water, and measured friction factors are compared with predictions of correlations presented earlier. Experimental results of friction factor and Poiseuille correlation are in the agreement for DI water for laminar flow. While the friction factor measurements for DI water exceed the predictions of Blasius correlation for turbulent region, they are in close agreement with Petukhov correlation.

Measured pressure drop for a Reynolds number range of 1400 to 4000 is shown in Fig. 5. Pressure drop raises with increasing particle fraction and flow rate, and the pressure drop increase of the nanofluids with respect to that of water is significantly higher in transitional flow when compared to laminar and turbulent regimes. Local zones referred as turbulent puffs begin to appear beyond laminar flow with increasing flow rate, and they lead to chaotic fluctuations in the pipe flow. Solid particles in the fluids lead to observable changes at the onset of the transitional flow, which is in agreement with [32,33]. The particle-particle interactions in nanofluids lead to a pressure drop increase in transition zone as suggested by [34]. Even though measured maximum pressure drop increase is 30% in transition regime, increase in the pressure drop is less than 10% for laminar and turbulent regimes. Therefore, operation within the transition regime should be further avoided to prevent high pressure drop when working with the graphene nanofluids.

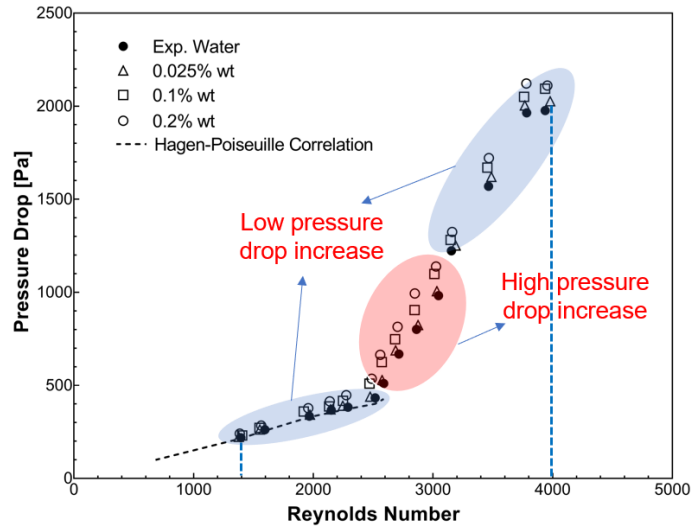

**Fig. 5.** Pressure drop change for different concentrations and flow rates

Transition from laminar to turbulent flow for water and graphene-water nanofluids are presented clearly in Fig. 6. Change in friction factor for water and graphene nanofluids are shown together with curves fitted on measured values. Transition region can be approximately identified from the intersection of the curves fitted to experimental data. For the laminar flow, friction factor declines with increasing flow rate, and the trends for water and graphene nanofluids are similar. Friction factor of DI water and nanofluid with 0.025% particle concentration is almost identical through the laminar region, whereas it is higher for the nanofluids with 0.1% and 0.2% at a given Reynolds number. For water and the nanofluid with 0.025% particle fraction, the increase in friction factor starts after a Reynolds number of ~2450 manifesting the onset of transition and continues up to ~3150, where further increase in Reynolds number induces to a decline in friction factor manifesting the onset of turbulent flow. It can be seen in Fig. 6 that transition starts at smaller Reynolds numbers for nanofluids with higher particle concentrations, in agreement with literature [23,35–38]. Furthermore, Matas et al. [34] defined transition threshold by means of particle size (d) and pipe diameter (D) in their experimental study. They proposed that if the D/d ratio is lower than 65, the transition shifts to the lower Reynolds numbers. It was shown in our previous study that the mean graphene nanoparticle size in the nanofluid is about 600 nm [27]; considering that the diameter of the test pipe used in this study is 6 mm, the transition shift we observed is consistent with their proposal.

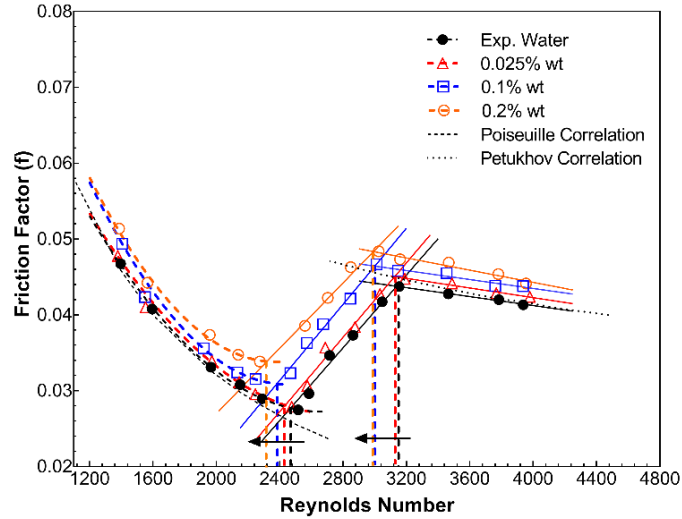

**Fig. 6**. Friction factor change for laminar, transition and turbulent flow

The predicted onset of laminar to turbulence transition and turbulent flow based on the approach explained above are listed in Table 1 for all concentrations by considering measurement uncertainty. The onset of transition shifts to a lower Reynolds number by ~4% for 0.1% particle mass fraction and ~7% for 0.2% particle mass fraction with transition phenomenon observed at lower Reynolds with increasing particle concentration. The early transition appears to be due to extra disturbance caused by the graphene nanoparticles in denser nanofluids as suggested in [39]. Beyond $Re_D = 3200$, friction factors for all samples decreases with increasing Reynolds number with trends similar to those by Eqs. 6 and 7 as the flow in the pipe becomes turbulent.

**Table 1.** Approximate onset of the transition and turbulence regimes in terms of Reynolds number according to pressure drop measurement

| Particle Concentration | Onset of Transition ($Re_D$) | Onset of Turbulence ($Re_D$) |
|---|---|---|
| 0 (pure water) | 2475 ± 50 | 3150 ± 65 |
| 0.025% | 2435 ± 50 | 3125 ± 65 |
| 0.1% | 2385 ± 50 | 3010 ± 60 |
| 0.2% | 2315 ± 45 | 2990 ± 60 |

**3.2- Heat Transfer Coefficient and Nusselt Number**

Similarly, convective heat transfer performance of graphene-water nanofluids is also investigated for nanofluids with 0.025, 0.1 and 0.2% particle mass fractions focusing on laminar to turbulent transition. Validation studies are carried out for DI water before testing nanofluids, by comparing measurements with predictions of Shah and London [28] and Gnielinski [29] correlations for laminar regime, and Gnielinski correlation [30] for turbulent regime. Experimental measurements are in the agreement with predictions based on correlations as seen in Fig. 7 and Fig. 8. It should be noted that the laminar regime in this study is hydrodynamically developed and thermally developing, whereas the flow is fully developed for most of the test section for turbulent flow.

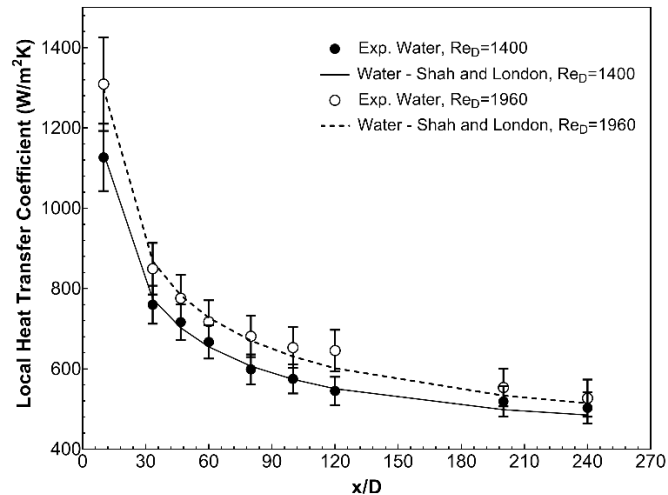

**Fig. 7.** Comparison of local heat transfer coefficient measurements with Shah and London correlation for DI water at $Re_D = 1400$ and $1960$

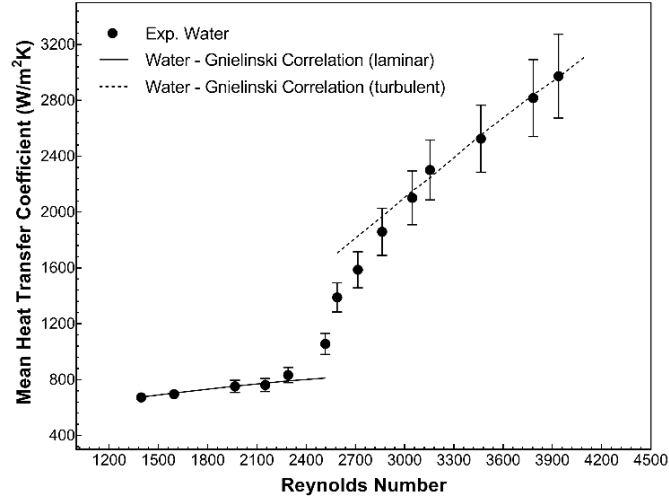

**Fig. 8.** Comparison of mean heat transfer coefficient measurements with Gnielinski correlations for DI water

Experiments are performed for a range of Reynolds numbers with three different nanoparticle concentrations to investigate effects of flow rate and particle concentration on heat transfer in different flow regimes. Local heat transfer coefficient of graphene-water nanofluids is measured at various flow rates for laminar flow and measurements are shown for $Re_D = 1400$ and $1950$ ($\pm 50$) in Fig. 9. Precise control of volumetric flow rate is not possible beyond a limit as a manual valve is used. Hence, the Reynolds numbers of water and nanofluids in Figs. 9 and 10 slightly differs, where the variance is within the measurement uncertainty. As seen in Fig. 9, local heat transfer coefficient, $h_x$, of both water and graphene nanofluids increases with flow rate, and particle concentration. The increase in the local heat transfer coefficient for different particle mass fractions at various axial locations and Reynolds numbers is shown in Table 2. The observed increase in heat transfer coefficient is similar for each concentration at $x/D=120$ and $x/D=240$ considering measurement uncertainty, and local heat transfer coefficient enhancement depends more significantly on the particle concentration in laminar flow rather than flow rate. Mean heat transfer coefficient, $\bar{h}_D$, is also investigated at a Reynolds number of 1400 and 1950 for better understanding the effective heat transfer mechanisms under laminar flow. Mean heat transfer coefficient enhancement for 0.025, 0.1 and 0.2% particle mass fraction is 7.3, 17.2 and 22.7% at $Re_D = 1400$ and 7.2, 17.6 and 22.8% at $Re_D = 1950$, respectively; which are similar to the increase in thermal conductivity (Fig. 2). Therefore, convective heat transfer coefficient increase is largely

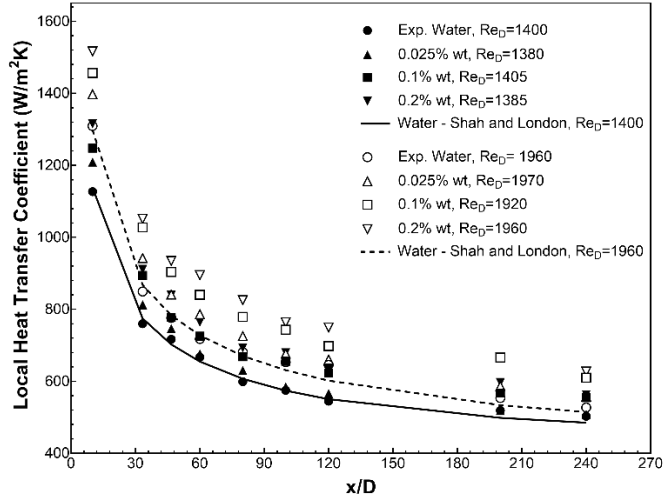

**Fig. 9.** Local heat transfer coefficient of graphene-water nanofluids for 0.025, 0.1 and 0.2% mass concentration at around $Re_D$=1400 and $Re_D$=1950

because of the thermal conductivity augmentation for laminar flow, which can be seen more clearly in Fig. 10. Unlike the case with the local heat transfer coefficient, local Nusselt number values do not change significantly for different concentrations, but change with different flow rates are consistent with predictions of Eq. 13. It can also be observed from Fig. 10 that the flow is hydrodynamically fully developed after *x/D*=120, but it is still thermally developing by the end of test unit.

**Table 2.** Local heat transfer coefficient increase of graphene-water nanofluids for $Re_D$=1400 and 1950 at *x/D*=120 and *x/D*=240

| For *x/D*=120 | | | |
|---|---|---|---|
| | **0.025 wt%** | **0.1 wt%** | **0.2 wt%** |
| $Re_D$=1400 | 3.6% | 12.1% | 15.4% |
| $Re_D$=1950 | 2.5% | 8.1% | 15.2% |
| For *x/D*=240 | | | |
| | **0.025 wt%** | **0.1 wt%** | **0.2 wt%** |
| $Re_D$=1400 | 3% | 11.2% | 14.8% |
| $Re_D$=1950 | 3.6% | 12.8% | 17.4% |

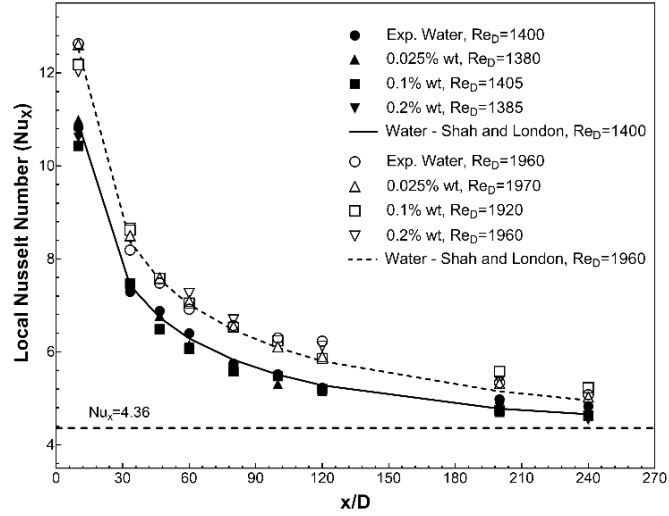

**Fig. 10.** Local Nusselt number of graphene-water nanofluids for 0.025, 0.1 and 0.2% mass fraction at around $Re_D=1400$ and $Re_D=1950$

Heat transfer in laminar to turbulence transition is investigated next, by further increasing the Reynolds number from 1950 to 4000. Fig. 11 illustrates the mean heat transfer coefficient, $\bar{h}_D$ change of graphene-water nanofluids with Reynolds number. In addition, mean heat transfer coefficient change for laminar, transition and turbulent flow regimes are identified from Fig. 11 using identical methodology with Fig. 6. For laminar flow, mean heat transfer coefficients of all three nanofluids show parallel trends, slightly increasing with Reynolds numbers nearly from 1400 to 2200 due to changing entry length. Whereas, the heat transfer behavior of water and graphene-water nanofluids starts to change beyond $Re_D = 2200$. The observed onset of the laminar to turbulent transition and turbulent flow are listed in Table 3.

Similar to the observations for the pressure drop measurements, transition starts at lower Reynolds numbers with increasing particle concentration. Dramatic enhancement in mean heat transfer coefficient of DI water at around a Reynolds number of 2450 is observed, and mean heat transfer coefficient converges to predictions of Gnielinski correlation nearly at 3150. While the nanofluid transition starts at lower flow rates than it does for water in the agreement with the studies mentioned earlier, the onset of transition observed relying on the heat transfer measurements correspond to even lower Reynolds numbers for tested nanofluids when compared to those based on the friction factor measurements. The onset of the transition for the nanofluids with 0.025, 0.1 and 0.2% particle mass concentrations are around Reynolds numbers of 2370, 2250 and 2200 considering heat transfer measurements, whereas they are 2435, 2385 and 2315, considering friction factor measurements. On the other hand, the onset of the turbulent flow is almost same for heat transfer measurements as accordance with friction factor measurements. The deviation between onset of transition based on the heat transfer coefficient and friction factor measurements being different can be attributed to the fact that the flow is not thermally fully developed at the exit of test unit as seen in Fig. 10. Therefore, the predictions based on friction factor constitute a more realistic inference for onset of transition. The lowering of onset of transition Reynolds numbers can be explained based on the introduced disturbance to the flow due to addition of the particles, causing micro-turbulence that helps inertia forces become more effective, leading to induced fluctuations at lower flow rates.

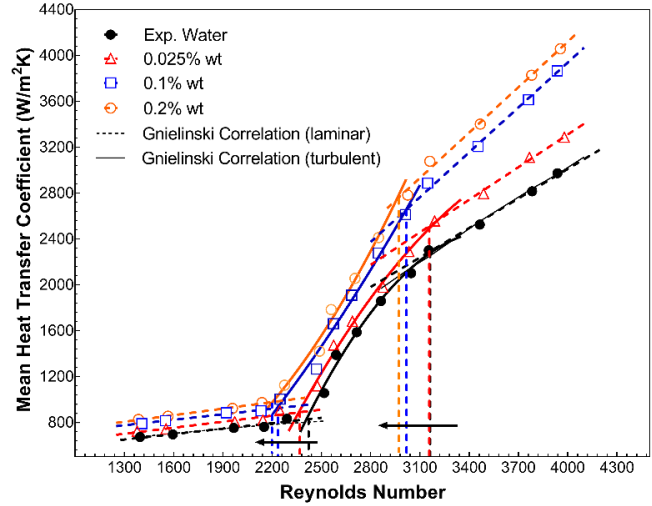

**Fig. 11.** Mean heat transfer coefficient change with Reynolds number for graphene-water nanofluids with 0.025, 0.1 and 0.2% mass concentration

**Table 3.** Onset of the transition and turbulent flow regimes in terms of Reynolds number according to heat transfer measurements

| Particle Concentration | Onset of Transition ($Re_D$) | Onset of Turbulence ($Re_D$) |
|---|---|---|
| 0 (pure water) | 2440 ± 50 | 3160 ± 65 |
| 0.025% | 2370 ± 50 | 3155 ± 65 |
| 0.1% | 2250 ± 45 | 3020 ± 60 |
| 0.2% | 2200 ± 45 | 2970 ± 60 |

While the thermal conductivity enhancement mechanisms are dominant for laminar flow, it appears that these mechanisms are not dominant after transition regime, especially for the higher concentrations. The mean Nusselt number change with Reynolds number is indicated in Fig. 12, where it is seen that Nusselt numbers of the nanofluids are very close to each other for the laminar flow. Whereas, the mean Nusselt numbers raise for higher particle fractions, with increasing Reynolds numbers. Mean heat transfer coefficient increase for the nanofluid with 0.025% concentration becomes more pronounced beyond a Reynolds number of 3100 and greatest increase is 11% at $Re_D = 3200$. Besides, the onset of turbulence for the nanofluids with 0.1 and 0.2% particle mass fractions is around $Re_D = 3000$. The maximum augmentation in the mean heat transfer coefficient is 30% and 36% for the nanofluids with 0.1 and 0.2% particle mass fraction at a Reynolds number of 3950, respectively. These results are in agreement with the literature [22,40].

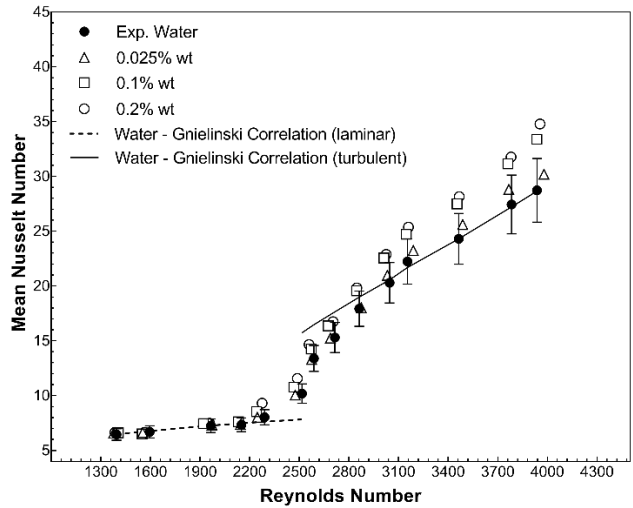

**Fig. 12**. Mean Nusselt number change of graphene-water nanofluids for 0.025, 0.1 and 0.2% mass concentration

It is stated in [27] that graphene nanoparticles create percolating chains in higher concentration nanofluids that constitutes the dominant thermal conductivity enhancement mechanism of suspension. Although the concentrations of the nanofluids investigated in this research are lower than the percolation threshold ($\phi \approx 1$ wt%), some percolating structures form as shown in Fig. 1b and these structures appear to be effective during the circulation in lower flow rates, creating localized zones of enhanced heat diffusion, leading to heat transfer augmentation.

Beyond laminar region, instabilities and random fluctuations in the flow disrupt the percolation structures, improving the dispersion of particles. While some researchers argued that mechanisms such as Brownian motion and thermophoresis do not have significant effect on heat transfer, some showed they might be responsible for the heat transfer increase by considering nanoparticle migration [41–45].

A scale analysis is performed considering the diffusion coefficients regarding Brownian diffusion $\left(D_B \frac{\Delta\varphi}{D}\right)$ and thermophoresis $\left(D_T \frac{\Delta T}{TD}\right)$, for identifying the effect of these two mechanisms [43]. Here,

$$D_B = \frac{k_B T}{3\pi\mu d_{np}} \tag{24}$$

$$D_T = \beta \frac{\mu}{\rho} \varphi \tag{25}$$

$$\beta = 0.26 \frac{k}{2k + k_{np}} \tag{26}$$

The scales of the terms used in the equations are $k_B \sim 10^{-23}$ J/K, $T \sim 10^2$ K, $\mu \sim 10^{-3}$ Pa.s, $d_{np} \sim 10^{-7}$ m, $\Delta\varphi \sim 10^{-3}$, $D \sim 10^{-3}$ m, $k_{nf} \sim 1$ W/m.K, $k_{np} \sim 10^3$ W/m.K, $\rho \sim 10^3$ kg/m$^3$, $\varphi \sim 10^{-1}$, $\Delta T \sim 10$. It is observed that the scale of Brownian diffusion coefficient ($D_B$) is two orders of magnitude smaller than that of thermophoretic diffusion coefficient ($D_T$). Therefore, thermophoresis is the more effective mechanism of heat transfer enhancement beyond laminar flow. This observation is consistent with that of Chandrasekar and Suresh [46], who argued that thermophoresis is more effective in the turbulent flow, where heat transfer increases much more than it does for laminar flow.

As a final note, viscosities of the nanofluid samples are measured before and after the experiments for investigating the possible effects of the experiments on graphene-water nanofluids. The measurements are performed in the temperature range of 25 to 50°C, repeated three times, and the average values are calculated. The maximum change in the relative viscosity before and after the tests is observed are less than ±2%. Hence, the rheological behavior of prepared nanofluids is not significantly affected from the experiments.

## 4- Conclusions

Forced convection in graphene-water nanofluids in a circular tube is experimentally investigated focusing on the transitional behavior. Experiments are carried out for a Reynolds number range of 1400 to 4000, and 0.025, 0.1 and 0.2% particle mass concentrations. The measured friction factor and heat transfer coefficient for water are compared to those calculated by correlations for validation of the test setup. Increase in the pressure drop for studied nanofluids does not exceed 10% in comparison with water in the laminar and turbulent regimes, whereas it is up to the 30% for the transitional flow. Convective heat transfer performance of the graphene-water nanofluids is also investigated considering laminar, transition and early turbulent regions to understand the effective mechanisms on the convective heat transfer in all these flow regimes. For laminar flow, local heat transfer coefficient enhancement for nanofluids at different flow rates, and $x/D=120$ and $x/D=240$ roughly does not change with Reynolds number. Besides, all nanofluids have similar Nusselt number for given Reynolds numbers and axial locations, whereas mean heat transfer coefficient increase is around 7, 17 and 22% for the nanofluids with 0.025, 0.1 and 0.2% particle mass concentration, respectively. Considering that increase is similar to that of thermal conductivity, dominant thermal enhancement mechanism is percolation that is effective for thermal conduction. Then, it is observed that onset of transition from laminar flow shifts to lower Reynolds numbers as particle concentration increases. Significant heat transfer coefficient enhancement is seen for turbulent flow of nanofluids. A scale analysis showed that thermophoresis is the dominant mechanism, compared to Brownian motion. Prepared nanofluids exhibit maximum of 36% mean heat transfer coefficient enhancement for 0.2% particle mass fraction at a Reynolds number of 3950.

It is observed that graphene-water nanofluids provide a significant heat transfer enhancement by creating percolating structures, and due to thermophoresis. Pressure drop increase is relatively low in the laminar and turbulent regimes, whereas it is higher in the transitional flow. Therefore, operation under transitional flow must be avoided for graphene-water nanofluids as it would limit the performance and efficiency due to resulting high pressure drop and pumping power.

**Acknowledgments**

The authors would like to thank Boğaziçi University for support under B.U. Research Fund AP-15302.